\documentclass[referee]{jfm}
\usepackage[T1]{fontenc}
\usepackage{graphicx}
\usepackage{amsmath}
\usepackage{bm}
\usepackage{caption}
\usepackage{subcaption}

\usepackage{cleveref}

\newcommand{\dtot}[2]{\frac{\mathrm{d}{#1}}{\mathrm{d}{#2}}}
\newcommand{\dpar}[2]{\frac{\partial {#1}}{\partial {#2}}}

\newcommand{\dd}{\mathrm{d}}

\newcommand{\nn}{\hat{\bm{n}}}
\newcommand\Ca{\mbox{\textit{Ca}}}  
\newcommand\We{\mbox{\textit{We}}}  

\shortauthor
{
  
  \'E. Ruiz-Guti\'errez,
  C. Semprebon,
  G. McHale and
  R. Ledesma-Aguilar
}

\title[Liquid Barrels in Wedges]
{
  Statics and Dynamics of Liquid Barrels in Wedge Geometries
}

\author
{
  \'Elfego Ruiz-Guti\'errez\aff{1},
  Ciro Semprebon\aff{1},
  Glen McHale\aff{1} \and
  Rodrigo Ledesma-Aguilar\aff{1} \corresp{\email{rodrigo.ledesma@northumbria.ac.uk}}
}

\affiliation
{
  \aff{1} Smart Materials \& Surfaces Laboratory,
  Department of Mathematics, Physics and Electrical Engineering,
  Ellison Place, Northumbria University, Newcastle upon Tyne, NE1 8ST, UK
}

\usepackage{epstopdf}

\begin{document}

\maketitle

\begin{abstract}

We present a theoretical study of the statics and dynamics of a partially wetting liquid droplet, of equilibrium contact angle $\theta_{\rm e}$, confined in a solid wedge geometry of opening angle $\beta$. 
We focus on a mostly non-wetting regime, given by the condition $\theta_{\rm e} - \beta > 90^\circ$, where the droplet forms a {\it liquid barrel} -- a closed shape of positive mean curvature. 
Using a quasi-equilibrium assumption for the shape of the liquid-gas interface, we compute the surface energy landscapes experienced by the liquid upon translations along the symmetry plane of the wedge.
Close to equilibrium, our model is in good agreement with numerical calculations of the surface energy minimisation subject to a constrained position of the centre of mass of the liquid. 
Beyond the statics, we put forward a Lagrangian description for the droplet dynamics. We focus on the the over-damped limit, where the driving capillary force is balanced by the frictional forces arising from the bulk hydrodynamics, the corner flow near the contact lines and the contact line friction. 
Our results provide a theoretical framework to describe the motion of partially wetting liquids in confinement, and can be used to gain further understanding on the relative importance of dissipative processes that span from microscopic to macroscopic length scales. 

\end{abstract}

\section{Introduction}
\label{section:Introduction}

The statics and dynamics of liquid droplets in wedge geometries is an active research topic across disciplines, spanning biological physics~\citep{prakash2008ratchet}, granular media~\citep{bocquet2002humid,kohonen2004granular,grof2008strength} and microfluidics~\citep{dangla2013droplet,renvoise2009drop,luo2014separation}.  More fundamentally, understanding the motion of droplets in wedges can shed light on complex phenomena, such as interfacial instabilities~\citep{Stone2012,Bico2016} and the impact of surface roughness on contact-line dynamics~\citep{moulinet2002roughness}. 

When a liquid droplet is brought into contact with the inner walls of a wedge-shaped channel, the system will tend to minimise its total surface energy. In general, the transient dynamics and the final equilibrium state can be characterised in terms of two main parameters, corresponding to the opening angle of the wedge, $\beta$, which characterises the confinement geometry, and the equilibrium contact angle of the liquid with the solid, $\theta_{\rm e}$, which quantifies the wetting properties of the liquid.  

Broadly speaking, one can identify three qualitatively different regimes for the behaviour of droplets in wedges depending on the interplay between $\beta$ and $\theta_{\rm e}$. The first corresponds to an `apex contact' regime, where $0^\circ \leq \theta_{\rm e}  \leq 90^\circ + \beta$. In such cases the liquid-gas interface is concave and forms a transient {\it capillary bridge} when placed between the walls of a wedge. It was first noted by \citet{Hauksbee1710} that the free motion of such structures (i.e., in the absence of external forces, such as gravity) always results in their migration towards the apex of the wedge. For situations where $0 \leq \theta_{\rm e}  \leq 90^\circ - \beta$, \citet{concus1969behavior}, and \citet{concus2001bridges} showed that a global equilibrium is not possible, leading to the complete spreading of the liquid along the wedge apex. 
On the other hand, when $90^\circ - \beta < \theta_{\rm e} \leq 90^\circ +\beta$, the liquid-gas interface forms an equilibrium shape that touches the apex of the wedge keeping a contact line of finite length, sometimes referred to as an `edge-blob'~\citep{concus1998discontinuous,concus2001bridges,brinkmann2004}.  
Recently, \citet{reyssat2014drops} studied the motion of completely-wetting capillary bridges within wedge-shaped channels and identified two regimes in the dynamics of the liquid.  Close to the apex, the main source of energy dissipation is the viscous friction in the bulk of the liquid, which balances the rate of work done by capillary forces. As a result, the time evolution of the position of the capillary bridge is linear. This picture changes when the liquid is far from the apex of the wedge, where the main source of dissipation is the corner flow near the apparent contact lines. The result is a different equation of motion, which is given by a power-law dependence of the position of the liquid as a function of time with an exponent $4/13$. 
 
A second regime corresponds to the reverse limiting wetting situation, where $\theta_{\rm e}=180^\circ$, and for which a liquid in a wedge-shaped channel will form a suspended droplet, a situation also found for gas bubbles. In such a case, a confined droplet  will always migrate away from the apex of the wedge~\citep{dangla2013droplet}. In sharp contrast to the complete-wetting limit, the equilibrium shapes of suspended droplets or bubbles  correspond to perfect spheres. The dynamics of such systems will often involve the interplay between the liquid/gas and the surrounding fluid~\citep{Bretherton1961,Park1984}. However, in the specific case of a low-viscosity fluid (air) suspended in a liquid of relatively high viscosity (silicone oil), \citet{reyssat2014drops} showed that the main sources of dissipation originate from the liquid, and that the same equations of motion that hold for completely wetting capillary bridges also hold for completely nonwetting bubbles. 

The third regime, which is the focus of this paper, corresponds to a mostly non-wetting situation, where $\theta_{\rm e} > 90^\circ + \beta$. In such a case, the liquid-gas interface is convex, i.e., it has a positive mean curvature. Therefore, upon contact with the walls of a wedge, a droplet will form a {\it liquid barrel}.  \citet{concus2001bridges} studied the equilibria of liquid barrels in wedge geometries. They showed that, in contrast to capillary bridges, liquid barrels form closed surfaces avoiding the apex of the wedge, and that, in the absence of external forces, such shapes correspond to sections of spheres. Experimentally, \citet{baratian2015shape} recently observed such equilibrium shapes using an electrowetting setup, and showed that a spherical equilibrium shape implies a vanishing net force acting on the liquid and that non-spherical static shapes appear when subjecting the liquid to the action of gravity.  

Whilst the equilibrium states of liquid barrels in a wedge geometry are now well understood, several questions regarding the statics and dynamics of these systems remain open. 
In particular, the statics and dynamics of non-spherical barrel shapes can only be understood by knowledge of the net restitutive capillary force (which can be inferred from the free-energy landscape), and of any resistive forces, such as a net external force or a friction force caused by the motion of the liquid. 
Importantly, understanding the motion of liquid barrels towards an equilibrium state can reveal details of dissipative processes at three different length scales, namely, the large-scale viscous friction caused by the bulk flow pattern, the viscous friction caused by the motion of the liquid near the contact line, often described as a corner flow, and the friction caused by the motion of the contact line itself. 

In the present article, we carry out a theoretical study of the statics and dynamics of a liquid drop that forms a barrel shape upon contact with the walls of a wedge-shaped channel. In \S\ref{section:Governing Equations} we introduce a near-equilibrium model for the morphology of the barrel and compute the corresponding free-energy landscapes as a function of the position 
of the barrel relative to the apex of the wedge. 
We compare our analytical results in the near-equilibrium limit to numerical computations of the surface energy using a minimisation algorithm that fixes a constraint in the centre of mass of the liquid. 
In \S\ref{section:Motion of the droplet near equilibrium} {we derive} the equations of motion of the liquid barrel {using a Lagrangian approach, and calculate the overall drag arising from the bulk, corner-flow and contact-line contributions to energy dissipation. Finally, in \S\ref{section:Discussion and Conclusions} we discuss the implications of our results.}


\section{Free-Energy Model}
\label{section:Governing Equations}

\begin{figure}
  \begin{center}
 \includegraphics[width=0.98\textwidth]{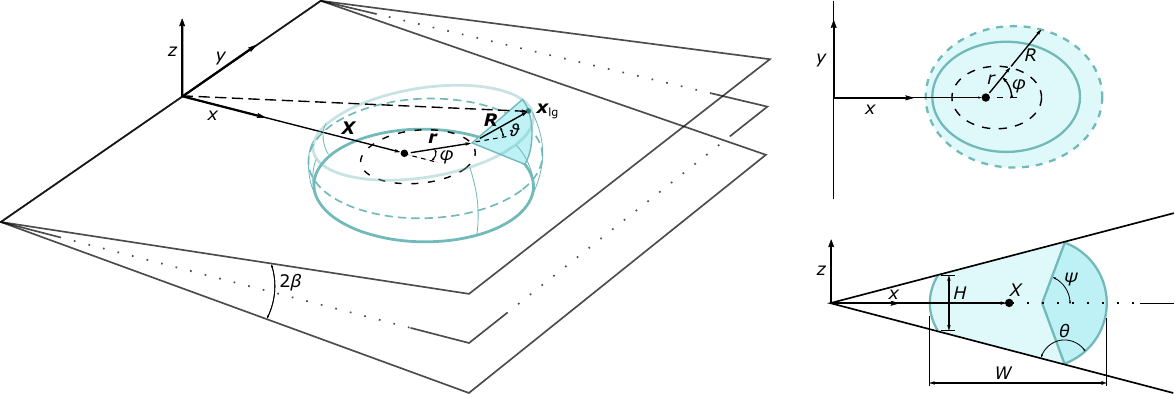}
  \end{center}
  \caption{ \label{fig:diagram_droplet_geometry} (Colour online) Schematics of the geometry of a liquid barrel inside a solid wedge of opening angle $2\beta$. The position vector of the liquid-gas interface, $\bm{x}_{\rm lg}$, is described using the vectors $\bm{X}$, $\bm{r}$ and $\bm{R}$, and the azimuthal and polar angles $\varphi$ and  $\vartheta$. The intersection with the solid, where  $\vartheta=\psi$, occurs at a prescribed contact angle $\theta$. The aspect ratio of the $xz$ cross section of the barrel is determined by its minimum thickness, $H$ and equatorial width, $W$.}
\end{figure}

Figure~\ref{fig:diagram_droplet_geometry} shows a schematic of the system under consideration, which consists of a liquid droplet that partially 
wets the inner surface of a wedge formed by two solid planes. 

We focus on a situation where the mass, $M$, temperature, $T$, and volume of the liquid, $V$, are held constant. The relevant thermodynamic potential is the Helmholtz free energy $\mathcal{F} = U - TS$, where $U$ and $S$ are the internal energy and entropy, respectively. 

From the second law of Thermodynamics, the Helmholtz free energy will either remain constant or decrease upon a change in the configuration of the system, i.e., $\delta \mathcal{F} \leq 0$. Such changes in the free energy are caused by the interfacial variations
\begin{equation}
  \dd F = \gamma \dd A_\mathrm{lg} + \gamma_\mathrm{sl} \dd A_\mathrm{sl} + \gamma_\mathrm{sg} \dd A_\mathrm{sg},
  \label{eq:helmholtz-free-general}
\end{equation}
where $\gamma$, $\gamma_\mathrm{sl}$, and $\gamma_\mathrm{sg}$ are the liquid-gas, solid-liquid, and solid-gas surface tensions respectively; and $A_\mathrm{lg}$, $A_\mathrm{sl}$, and $A_\mathrm{sg}$ are the corresponding  interfacial areas. Therefore, equilibrium states correspond to minima of the surface energy
\begin{equation}
  F = \gamma (A_\mathrm{lg} - A_\mathrm{sl} \cos \theta_\text{e}),
  \label{eq:free-energy}
\end{equation}
where the equilibrium angle, $\theta_\text{e}$, is determined by Young's Law,
\begin{equation}
  \cos \theta_\text{e} = \frac{\gamma_\mathrm{sg} - \gamma_\mathrm{sl}}{\gamma}.
  \label{eq:young-dupre}
\end{equation}

\subsection{Geometry}
\label{subsection:Geometry}

To determine $F$, we need to specify a suitable parametrisation of the geometry of the droplet,  as shown in Figure~\ref{fig:diagram_droplet_geometry}.  In Cartesian coordinates, the wedge walls are oriented at an angle $\beta$ from the $xy$ plane and intersect along the $y$ axis.  The unit normals to the walls are $\pm \nn(\pm \beta)$, where $\nn (\beta) = (- \sin \beta,\, 0,\, \cos \beta)$. 
We assume that the wedge walls are identical and perfectly uniform, implying a reflection symmetry about the bisector plane.

We describe a point on the liquid-gas interface using the position vector
\begin{equation}
  \bm{x}_{\mathrm{lg}} \equiv \bm{X} + \bm{r} + \bm{R}.
  \label{eq:xlg}
\end{equation}
The vector $\bm{X} = (X,\,0,\,0)$  defines the position of the geometric centre of the droplet, $X$, relative to the apex of the wedge. The vector $\bm{r} = r(\varphi) \, \hat{\bm{r}}$ is coplanar to the bisector plane, i.e., $\hat{\bm{r}} = (\cos \varphi,\, \sin \varphi,\, 0)$, where $\varphi$ is an azimuthal angle.  The vector $\bm{R} = R(\varphi,\vartheta) \, \hat{\bm{R}}$ points in the direction of the unit vector $\hat{\bm{R}} = (\cos \varphi \cos \vartheta,\, \sin \varphi \cos \vartheta,\, \sin \vartheta)$, where the polar angle $\vartheta$ subtends between the top and bottom walls. The combination of $r(\varphi)$ and $R(\varphi,\vartheta)$ can therefore be used to specify the shape of the liquid-gas interface.

Whilst the azimuthal angle varies in the interval $\varphi \in [0,\, 2\pi)$, the polar angle is restricted by the intersection of the liquid-gas interface with the solid walls, i.e., $\vartheta \in [-\psi, \psi]$, where the maximum angle, $\psi$, can be found by the geometrical condition  
\begin{equation}
  \nn(\pm \beta) \bcdot \bm{x}_\mathrm{lg}(\varphi, \vartheta=\pm \psi) = 0.
  \label{eq:contact-line}
\end{equation}
In addition, one can write a relation for the contact angle of the liquid-gas interface with the solid, $\theta$, measured from the liquid phase, which reads
\begin{equation}
  - \cos \theta = \pm \nn(\pm \beta) \bcdot \frac{\partial_\varphi \bm{x}_\mathrm{lg} \times \partial_\vartheta \bm{x}_\mathrm{lg}}{| \partial_\varphi \bm{x}_\mathrm{lg} \times \partial_\vartheta \bm{x}_\mathrm{lg} |} (\varphi, \vartheta = \pm \psi).
  \label{eq:gentheta}
\end{equation}

The aspect ratio of the droplet can be characterised by the height-to-width ratio,
\begin{equation}
  h \equiv \frac{H}{W},
  \label{eq:aspectratio}
\end{equation}
where the droplet height, $H \equiv |\bm{x}_\text{lg}(\upi, \psi) - \bm{x}_\text{lg}(\upi, -\psi)|,$ is the length of the line connecting the contact lines at the narrow end of the wedge and the droplet width, $W\equiv |\bm{x}_\text{lg} (0, 0) - \bm{x}_\text{lg} (\pi, 0)|$, is the distance between the leading and trailing points of the droplet's equator.

Using the parametrisation $\bm{x}_\mathrm{lg}(\varphi,\vartheta)$, we define the surface elements of the liquid-gas and solid-liquid interfaces as
\begin{eqnarray}
  \dd \bm{A}_\mathrm{lg} & \equiv & (\partial_\varphi \bm{x}_\mathrm{lg} \times \partial_\vartheta \bm{x}_\mathrm{lg}) \dd \vartheta \dd \varphi,
  \label{eq:dA_lg} \\
  \dd \bm{A}_{\mathrm{sl}} & \equiv & \frac{1}{2} \bm{x}_\text{lg}(\varphi, \psi) \times \frac{\dd }{\dd \varphi} \bm{x}_\text{lg}(\varphi, \psi) \, \dd \varphi.
  \label{eq:dA_sl}
\end{eqnarray}
Therefore, the surface energy $F$, and the volume of the droplet, $V$, can be expressed as
\begin{equation}
  F = \gamma \int_0^{2 \upi} \int_{-\psi}^{\psi} | \dd \bm{A}_\mathrm{lg} | - 2 \gamma \cos \theta_{\rm e} \int_0^{2 \upi} \nn  \bcdot \dd \bm{A}_\text{sl}, 
  \label{eq:freeenergygeneral}
\end{equation}
and 
\begin{equation}
  V = \frac{1}{3} \int_{\rm liq.} \bnabla \bcdot (x,y,z) \; \dd x \dd y \dd z = \frac{1}{3} \int_{0}^{2\upi} \int_{-\psi}^\psi \bm{x}_{\mathrm{lg}} \bcdot \dd \bm{A}_\text{lg},
  \label{eq:vgencart}
\end{equation}
where we have made use of the divergence theorem in the last equality.



\subsection{Morphology of liquid barrels}
\label{section:Morphology of the droplet}

For a liquid droplet of typical density $\rho \approx 10^3 \,\mathrm{kg}\,\mathrm{m}^{-3}$, dynamic viscosity $\eta \approx 1 \,\mathrm{mPa}\,\mathrm{s}$, surface tension $\gamma \approx 20-70 \,\mathrm{mN}\,\mathrm{m}^{-1}$ and characteristic linear size $V^{1/3} \approx 1 \,\mathrm{mm}$, the typical translation speed inside a narrow wedge, $\beta \sim1^\circ-10^\circ$, is $U\sim {\rm 1 - 10 ~mm~s^{-1}}$. Therefore, the Reynolds number $\Rey \equiv \rho U V^{1/3} \beta / \eta \sim 10^{-1}-10^0$, the capillary number, $\Ca \equiv \eta U / \gamma \sim 10^{-5} - 10^{-3}$, and the Weber number, $\We \equiv \Rey \Ca \sim 10^{-6}-10^{-3}.$

The smallness of the Weber and Reynolds numbers implies that perturbations to the liquid-gas interface of the droplet decay over a short timescale relative to the timescale of translational motion the drop~\citep{miller1968oscillations,zhongcan1987instability,landau2013fluid}, which we will describe by focusing on the regime of lubrication flow~\citep{Oron1997} by restricting our discussion to the limit of small wedge angles ($\beta \ll 1 $). Here, the smallness of the capillary number indicates that capillary forces dominate over the viscous bending of the 
interface, including the region near the contact lines~\citep{voinov1976hydrodynamics,cox1986dynamics1}. 

Therefore, we describe the near-equilibrium shape of the droplet as a smooth barrel shape intersecting the solid at the equilibrium contact angle. 
In terms of the parametrisation introduced in \S\ref{subsection:Geometry},  this corresponds to introducing the following approximations:
\begin{eqnarray}
  \theta & = & \theta_{\rm e}, \label{eq:shapetheta}\\
   R     & = & R(\varphi),\label{eq:shapeR}\\
   W / 2 & = &  R + r  = \text{const.} \label{eq:shapeW} 
\end{eqnarray}
where $R(\varphi)$ and $W/2$ are the local radii of curvature in the normal and tangent directions to the bisector plane, respectively.  
In terms of~(\ref{eq:shapetheta})--(\ref{eq:shapeW}), $X$ reduces to the average distance between the leading and trailing edges of the barrel's equator. 

\subsection{Equilibrium}
\label{subsection:Equilibrium}

We first analyse the equilibrium shapes of liquid barrels in wedge geometries. \citet{concus2001bridges} proved the existence of equilibrium states in the range $\theta_{\rm e} - \beta > 90^\circ$, corresponding to sections of spheres intersecting the walls of the wedge with the equilibrium angle $\theta_{\rm e}$. \citet{baratian2015shape} showed that for such solutions the surface tension acting on the wall integrated over the contact line exactly matches the pressure exerted by the liquid integrated over the solid-liquid interface.  

In terms of (\ref{eq:shapetheta})--(\ref{eq:shapeW}), such force-free spherical shapes can be recovered by setting $r = 0$ and $R=R_{\rm e}$, where 
\begin{eqnarray}
R_\text{e} & = & \left[ \frac{6 V}{\upi (\cos 3 \theta_\text{e} - 9 \cos \theta_\text{e})}\right]^{1/3}.
\label{eq:equilibrium-1-R}
\end{eqnarray}
This yields the following relations for the equilibrium position, $X_\text{e}$, height-to-width ratio, $h_\text{e}$, and surface energy, $F_\text{e}$:
\begin{eqnarray}
  X_\text{e} & = & -\frac{\cos \theta_\text{e}}{\sin \beta} R_{\rm e}, \label{eq:equilibrium-1-X}   \\
  h_\text{e} & = & - \cos (\theta_\text{e} - \beta),
  \label{eq:equilibrium-1-h} \\
  F_\text{e} & = & \gamma \frac{\upi}{3} (\cos 3 \theta_\text{e} - 9 \cos \theta_\text{e}) R_\text{e}^2.
  \label{eq:equilibrium-1-F}
\end{eqnarray}

\begin{figure}
\centering

\includegraphics[width=0.75\textwidth]{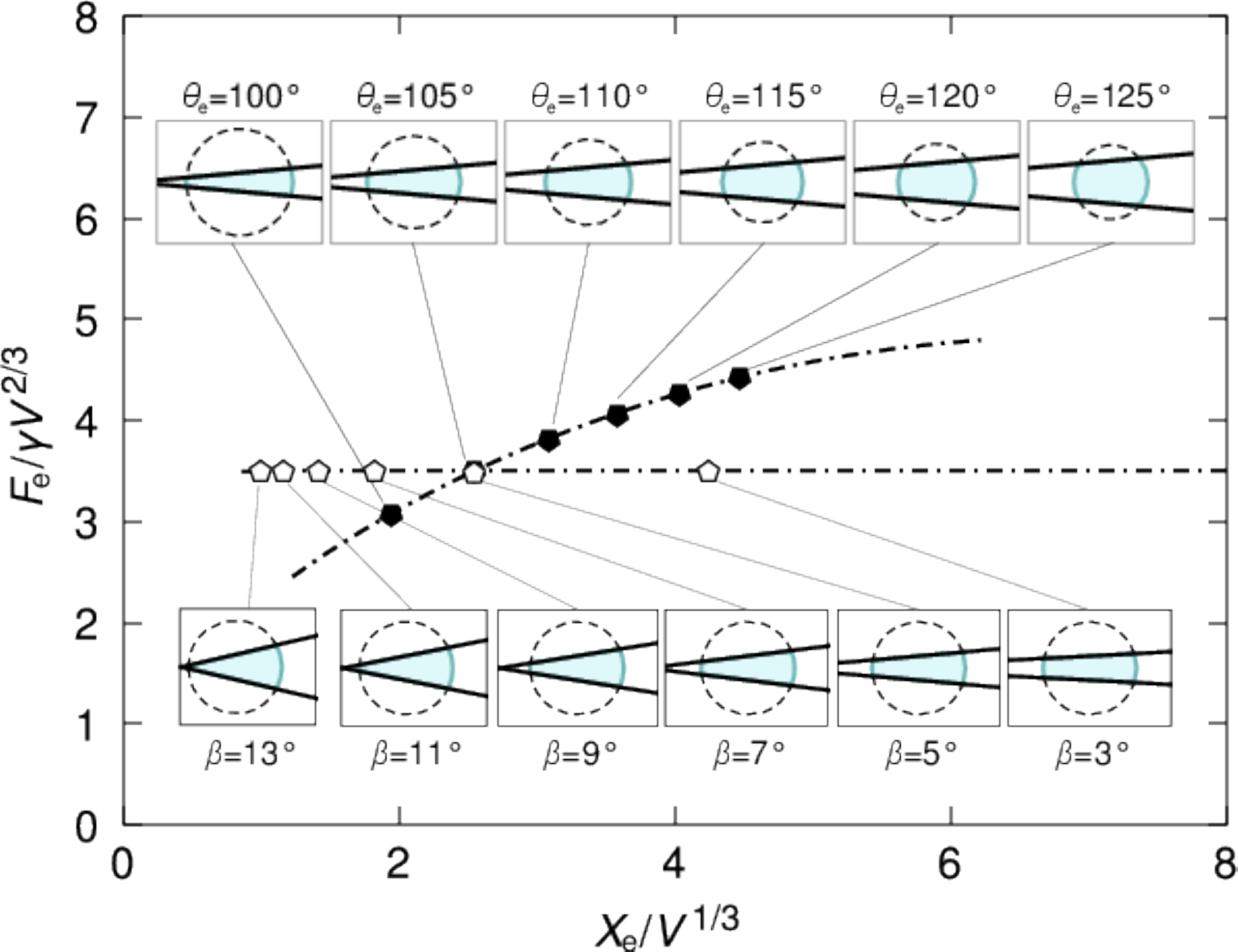}
\caption{\label{fig:equilibria} (Colour online) Equilibrium surface energy, $F_{\rm e}$, as a function of the distance from the wedge apex, $X_{\rm e}$, for different equilibrium contact angles (full symbols) and wedge angles (empty symbols). The insets correspond to cross sections of the barrels along the $xz$ plane.}
\end{figure}

Figure~\ref{fig:equilibria} shows the equilibrium surface energy of liquid barrels at different positions within the wedge. We first focus on the effect of $\theta_{\rm e}$ on $F_{\rm e}$ and $X_{\rm e}$. For $\theta_{\rm e}<180^\circ$, a suspended droplet will always reduce the total surface energy by wetting the walls of the wedge. This wetted area is be larger for smaller $\theta_{\rm e}$, and, because of volume conservation, the liquid settles at an equilibrium position closer to the wedge apex (see insets in figure~\ref{fig:equilibria}). 
At first sight, one might expect a similar effect by increasing the wedge angle, $\beta$. Indeed, from~(\ref{eq:equilibrium-1-X}) an increase in the wedge angle leads to 
a closer position of the barrel to the wedge apex. The surface energy, however, remains constant. Geometrically, this can be understood by noting that a change in $\beta$ is equivalent to a rotation of the solid walls about the centre of the sphere, which does not alter the size of any of the interfaces of the barrel, as shown in the insets of figure~\ref{fig:equilibria}.   

Note that for the droplet to form a closed barrel, that is, a structure that bridges the walls of the wedge avoiding its apex, one must have $R_\text{e} < X_\text{e}$, or, equivalently, 
\begin{equation}
h_\text{e}>0. 
\end{equation} 
From~(\ref{eq:equilibrium-1-h}), this condition is satisfied only if 
\begin{equation}
  \theta_\text{e} - \beta > 90^\circ.
  \label{eq:condition-for-equilibrium}
\end{equation}

Equilibrium states can also exist if $\theta_\text{e} - \beta \leq 90^\circ$ but not as liquid barrel shapes. In such cases it has been shown that the liquid completely invades the wedge~\citep{reyssat2014drops} and forms edge blobs~\citep{concus1998discontinuous,concus2001bridges} or filaments that spread laterally along the wedge apex~\citep{brinkmann2004}. 

For a parallel-plate geometry ($\beta = 0^\circ$), force-free barrels can exist provided that the separation between the solid walls matches the equilibrium height  
\begin{equation}
H_{\rm e} = 2 h_{\rm e} R_{\rm e},
\end{equation}
which follows easily from~(\ref{eq:equilibrium-1-h}). 
As noted by~\citet{kusumaatmaja2010bridges}, a displacement of the solid wall from this equilibrium configuration will still result in mechanical equilibrium, albeit in the presence of a
net external force. This situation can also occur for capillary bridges ($\theta_{\rm e} < 90^\circ - \beta $), for which no force-free equilibrium configurations can exist and the net force exerted by the liquid on the solid plates is always attractive.  

\subsection{Out-of-equilibrium shape and energy landscapes}
\label{subsection:Near equilibrium}

Out of equilibrium, the shape of the liquid barrel can be determined from (\ref{eq:xlg})--(\ref{eq:gentheta}), subject to (\ref{eq:shapetheta})--(\ref{eq:shapeW}). 
However, under these assumptions the unit vector $\hat{\bm{R}}$ is approximately normal to the contact line. Therefore, the boundary condition~(\ref{eq:gentheta}) can be replaced by the constraint 
\begin{equation}
  - \cos \theta_\text{e} = \nn(\beta) \bcdot  \hat{\bm{R}}(\vartheta = \psi).
  \label{eq:contactangleapprox}
\end{equation}
Combining these equations determines the transverse radius of curvature, 
\begin{eqnarray}
 R(\varphi) & = & q\left(1 + \epsilon \frac{   \cos \varphi}{\cos \varphi - \cos \theta_\text{e} / \sin \beta}\right), \label{eq:rq}
\end{eqnarray}
where 
\begin{equation}
  q \equiv - \frac{\sin \beta}{\cos \theta_\text{e}} X,
  \label{eq:defofq}
\end{equation}
and
\begin{equation}
 \epsilon \equiv \frac{W}{2 q} - 1.
  \label{eq:epsilon-def}
\end{equation}
Here, $q$ is a rescaled position of the geometric centre of the barrel, which reduces to the barrel's radius in equilibrium, i.e., $q (X_{\rm e}) = R_{\rm e}$. The parameter $\epsilon$, on the other hand, quantifies displacements of the equatorial radius of curvature from the spherical configuration.  

For a given volume, choosing the position of the barrel fixes its equatorial width, and consequently, $q$ and $\epsilon$ are not independent.  Rather, evaluating (\ref{eq:vgencart}) gives  the relation
\begin{equation}
  V(q, \epsilon) = q^3 \sum_{i=0}^{3} a_i \epsilon^i,
  \label{eq:vpoly}
\end{equation}
where the constants $a_i$ are functions of $\beta$ and $\theta_{\rm e}$, and are reported in Appendix A. Their expressions, however, simplify considerably in the limit of small wedge angles. Therefore, 
\begin{eqnarray}
  a_0 & = & \frac{\upi}{6} (\cos 3 \theta_\text{e} - 9 \cos \theta_\text{e}), \\
  a_1 & = & \upi (2 \theta_\text{e} - \upi - \sin 2 \theta_\text{e}) + O(\beta^2), \\
  a_2 & = & - 2 \upi \cos \theta_\text{e} + O(\beta^2), \\
  a_3 & = & 0 + O(\beta^2).
  \label{eq:vol_coefficients}
\end{eqnarray}
Using this approximation, and inverting (\ref{eq:vpoly}), we find 
\begin{equation}
  \epsilon(q) = \frac{1}{2 a_2} \left( \left\{ a_1^2 + 4 a_2 \left( \frac{V}{q^3} - a_0 \right) \right\}^{1/2} - a_1 \right).
  \label{eq:epspath}
\end{equation}

In order to evaluate the free energy (\ref{eq:freeenergygeneral}), we first express $F$ as a polynomial expansion in $\epsilon$. After some manipulations, we obtain  
\begin{equation}
  F(q, \epsilon) = \gamma q^2 \sum_{i=0}^3 (3 - i) a_i \epsilon^i + O(\epsilon^4).
  \label{eq:epoly}
\end{equation}
The constant-volume energy landscapes, $F_V(X)$, are then obtained by inserting (\ref{eq:epspath}) into (\ref{eq:epoly}) and recovering the definition of $q$ from (\ref{eq:defofq}), i.e., 
\begin{equation}
F_V(X) \equiv F \circ \epsilon \circ q (X). 
  \label{eq:energycomposition}
\end{equation}
Figure~\ref{fig:freeEnergyLandscape}(a) shows the energy landscapes for several values of $\theta_{\rm e}$ but keeping $\beta=5^\circ$. The asymmetry in the landscapes about the equilibrium position arises from the intrinsic asymmetry of the geometry of the wedge. A displacement towards the apex of the wedge induces a comparatively larger increase in the solid-liquid surface area relative to the liquid-gas surface area, and results in a sharper increase in the surface energy.  This same feature is observed in figure~\ref{fig:freeEnergyLandscape}(b), where we present energy landscapes at fixed $\theta_{\rm e} = 105^\circ$ and different values of $\beta$.

\subsection{Pressure profiles \& comparison to constrained surface-energy minimisation}
\begin{figure}
  \centering
  \begin{subfigure}{0.48\textwidth}
 \includegraphics[width=\textwidth]{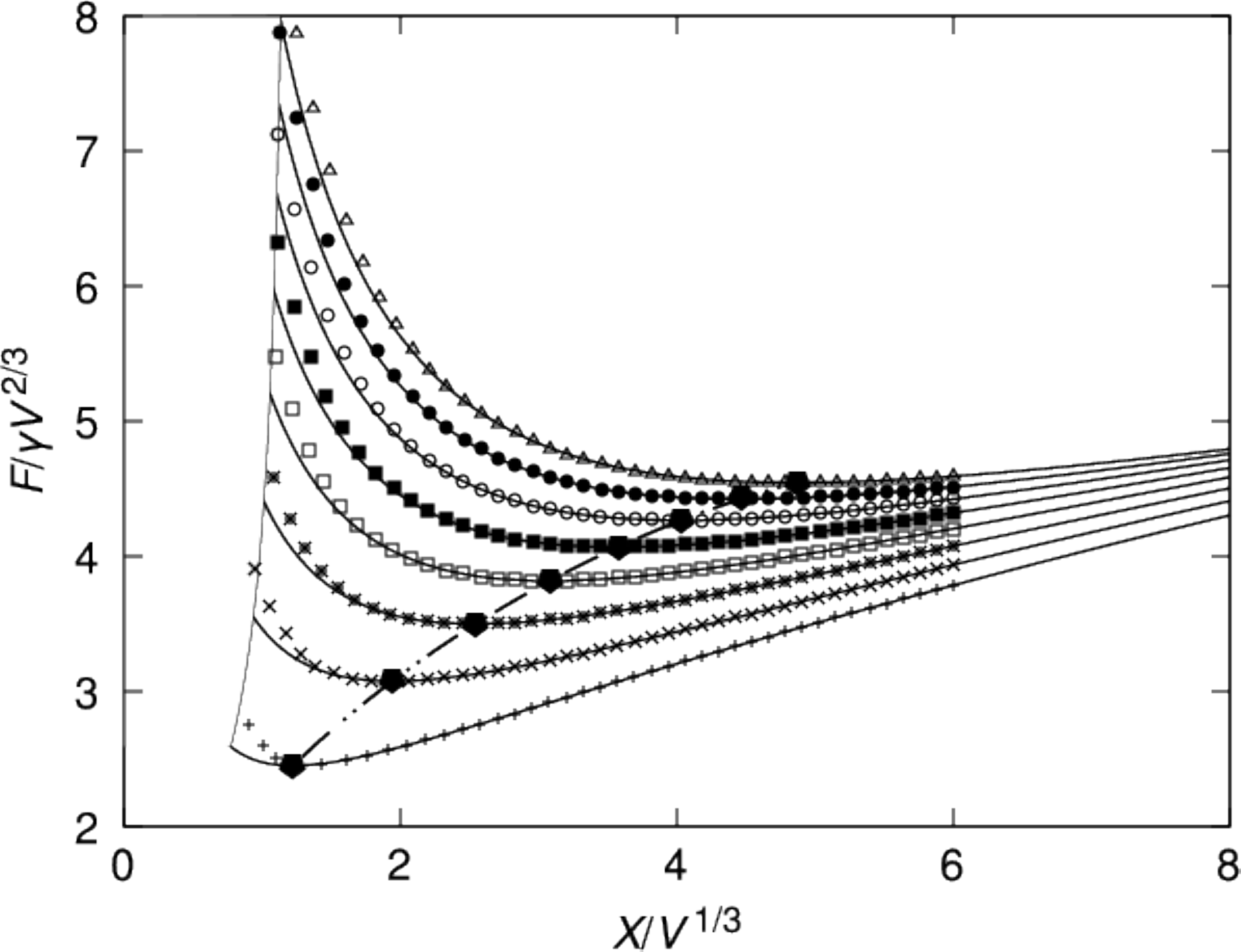}
    \caption{ \label{sfig:F_V-X-th} }
  \end{subfigure}
  ~
  \begin{subfigure}{0.48\textwidth}
    \includegraphics[width=\textwidth]{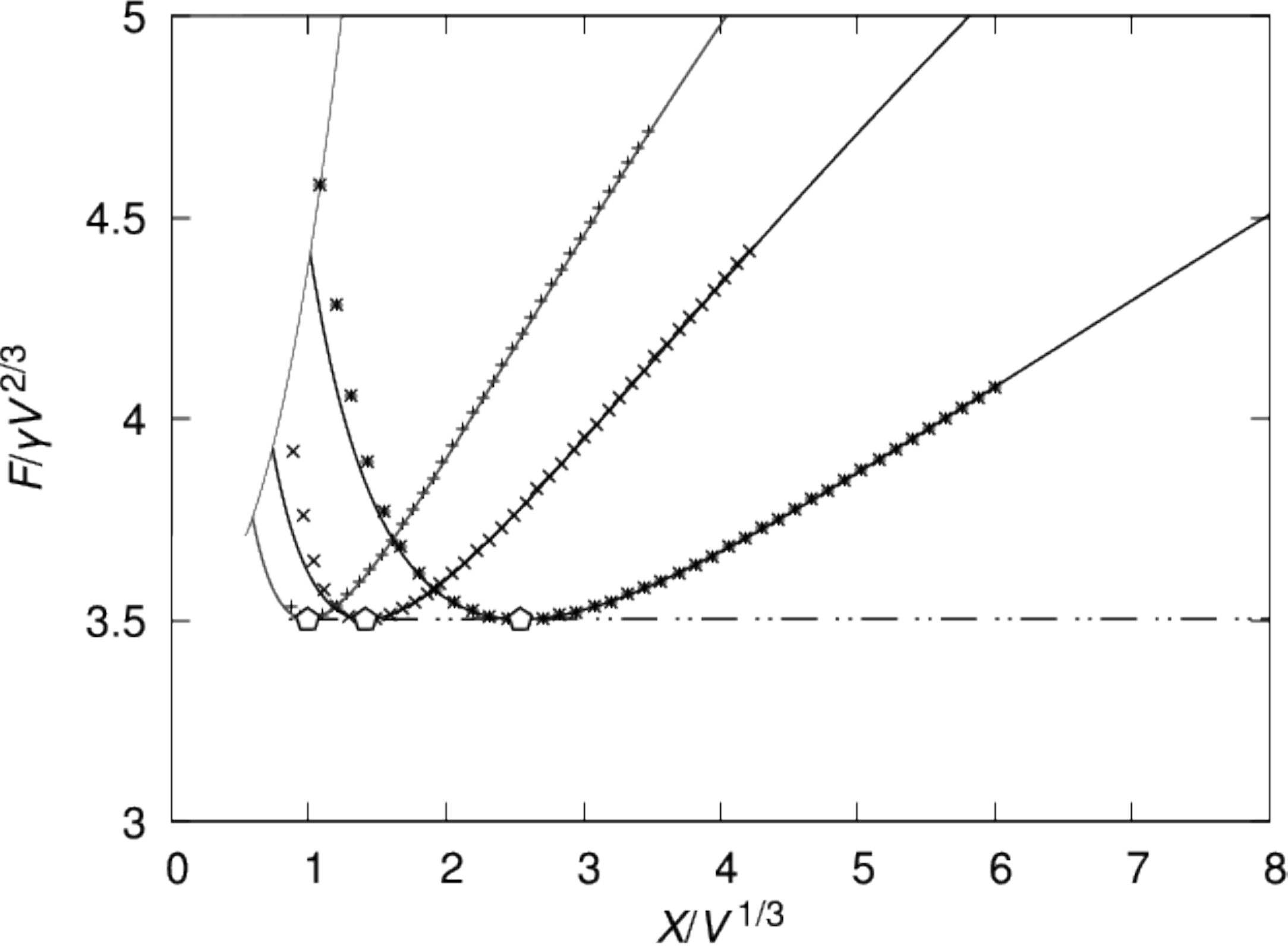}
    \caption{ \label{sfig:F_V-X-bt} }
  \end{subfigure}
  \caption{ \label{fig:freeEnergyLandscape} Energy landscapes along the position of the liquid barrel within the wedge, $X$, calculated analytically (solid lines) and numerically {\it via} constrained minimisation of the surface energy (symbols). (a) Curves for fixed $\beta = 5^\circ$ and different equilibrium contact angles: $\theta_{\rm e} = 95^\circ$ (+), $\theta_{\rm e} = 100^\circ$ ($\times$), $\theta_{\rm e} = 105^\circ$ (+\llap{$\times$}), $\theta_{\rm e} = 110^\circ$ ($\square$), $\theta_{\rm e} = 115^\circ$ ($\blacksquare$), $\theta_{\rm e} = 120^\circ$ ($\circ$), $\theta_{\rm e} = 125^\circ$ ($\bullet$), and $\theta_{\rm e} = 130^\circ$ ($\triangle$).
(b) Curves for fixed $\theta_{\rm e} = 105^\circ$ and different wedge angles: $\beta=5^\circ$ (+), $\beta=9^\circ$ ($\times$) and $\beta=13^\circ$ (+\llap{$\times$}). The pentagons correspond to the minima in the analytical curves. The solid cut-off lines correspond to the limit where the liquid-gas interface touches the apex of the wedge.}
\end{figure}

\begin{figure}
  \centering

   \includegraphics[width=0.45\textwidth]{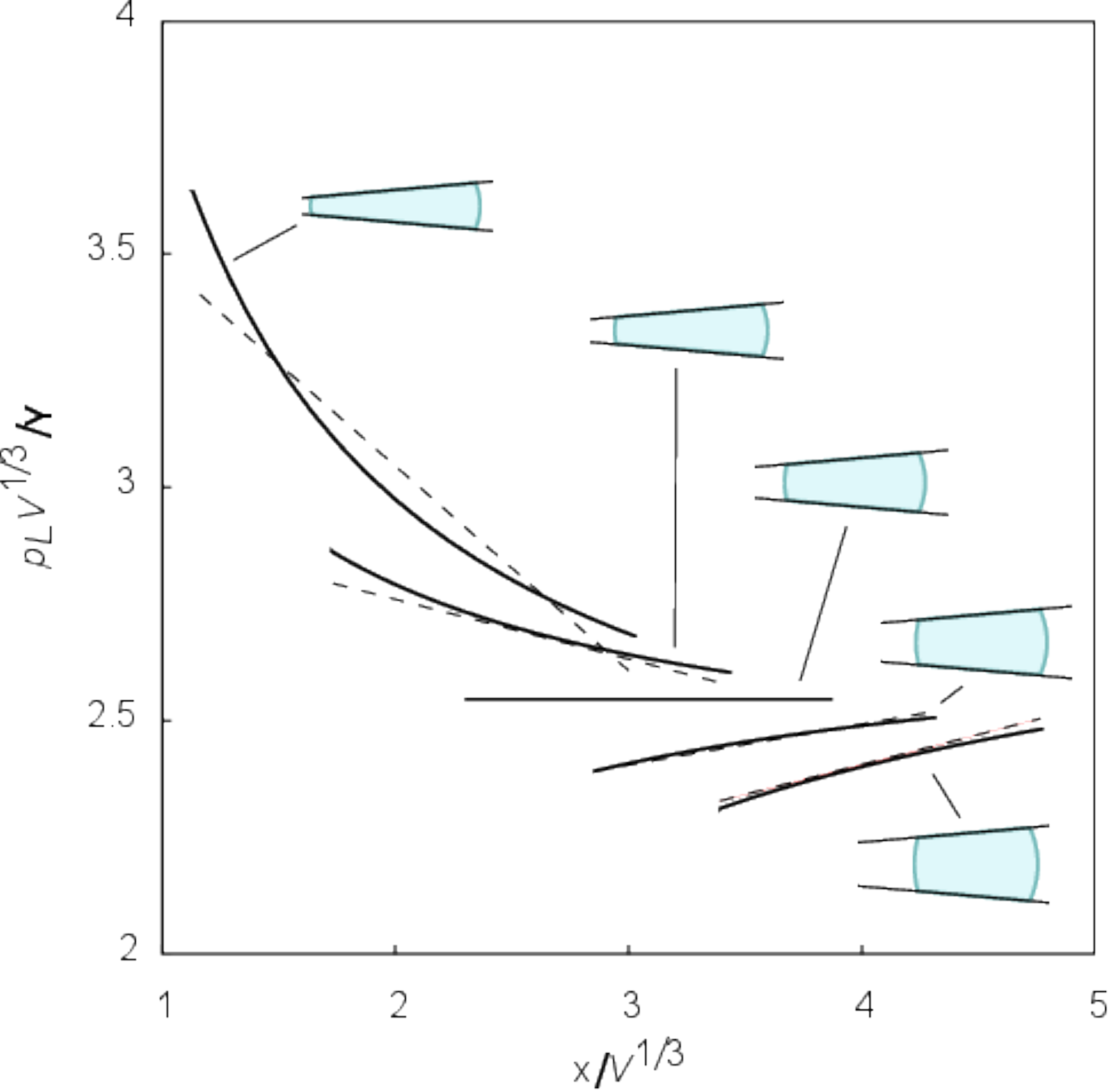}
  \caption{ \label{fig:LaplacePressure} (Colour online) Pressure profiles for out-of-equilibrium barrel shapes. 
  The continuous lines correspond to the Laplace pressure profile calculated directly from the analytical model. 
  The dashed lines correspond to hydrostatic pressure profiles obtained numerically from the constrained minimisation of the surface energy. Each pair of curves corresponds to a different displacement from equilibrium, from left to right, $(X-X_{\rm eq})/V^{1/3}=-1,\ -1/2,\ 0,\ 1/2,\ 1$.}
\end{figure}

To better understand the properties of the out-of-equilibrium barrel shapes presented in \S\ref{subsection:Near equilibrium}, we first focus on the Laplace pressure profiles arising from the local curvature of the liquid-gas interface,  $p_{\rm L} (x) = \gamma \left[2/W(x) + 1/R(x)\right]$.
Figure~\ref{fig:LaplacePressure} shows plots of $p_{\rm L}$ as a function of the bisector coordinate $x$ for barrels of equal volume and equilibrium contact angle, but different positions of the geometrical centre of the barrel $X$ along the wedge bisector. In equilibrium, the Laplace pressure profile across the droplet is uniform, and corresponds to $p_{\rm L}(x) =  2 \gamma/R_{\rm e}$. Out of equilibrium, $p_{\rm L}$  decreases along $x$ if $X<X_{\rm eq}$, and increases with $x$ if $X>X_{\rm eq}$. Notably, equal inwards and outwards displacements of the centre of the barrel, $X$,  lead to qualitatively different pressure profiles. For example, the right-most curve in figure~\ref{fig:LaplacePressure}, corresponding to $X-X_{\rm eq} = V^{1/3}$, shows a close to linear profile, whilst the left-most curve, where $X-X_{\rm eq} = -V^{1/3}$, mildly departs from linearity. 

The close to linear pressure profiles in figure~\ref{fig:LaplacePressure} suggest that, in such cases, the out-of-equilibrium barrel morphologies 
conform to the effect of a constant force density, such as gravity. To verify this hypothesis, we employed a finite element approach to numerically compute the barrel morphologies in mechanical equilibrium subject to a constraint in the position of the centre of mass. To this aim we used the public domain software SURFACE EVOLVER~\citep{Brakke1992} to define a triangulated mesh describing the liquid 
surface and to minimise the surface energy through a conjugate gradient algorithm. 

In the numerical method, the Lagrange multiplier of the volume constraint, $\lambda_V$, plays the role of the Laplace pressure at the coordinate $x=0$, while the Lagrange multiplier of the centre of mass, $\lambda_X$, can be interpreted as an effective body force required to hold the droplet in place. Therefore, a linear hydrostatic pressure profile can be constructed by writing $P(x)=\lambda_V+\lambda_X V^{2/3}x/\gamma $. In figure~\ref{fig:LaplacePressure} we overlay the linear pressure profiles obtained numerically to the analytical curves. The range of each curve  corresponds to the equatorial width of the barrel in each model.  As expected, there is good agreement in the magnitude of the pressure and in the location of the edges close to equilibrium with the analytical model, particularly for for $X>X_{\rm e}$. This agreement is also observed when comparing the analytical and numerical energy landscapes,  as shown in figures~\ref{fig:freeEnergyLandscape}(a) and~\ref{fig:freeEnergyLandscape}(b).  

\section{Motion of the droplet near equilibrium}
\label{section:Motion of the droplet near equilibrium}

\subsection{Lagrangian formulation}
The out-of-equilibrium barrel shape, given by (\ref{eq:rq}), is controlled by the coordinates $q$ and $\epsilon$. 
Here we consider the Lagrangian $L(q,\dot q, \epsilon, \dot \epsilon)$, where $q(t)$ and $\epsilon(t)$ are treated as dynamical variables and $\dot q\equiv \dd q/\dd t$ and $\dot \epsilon \equiv \dd \epsilon/\dd t$ are the corresponding velocities.  As discussed in \S\ref{section:Morphology of the droplet}, we focus on the overdamped regime, where inertial effects are negligible. Therefore, $L$ can be written purely in terms of the surface energy, $F$. Imposing the constraint of a constant volume, $V = V_0$, gives
\begin{equation}
  L = - F + p (V - V_0),
  \label{eq:lagrangianform}
\end{equation}
where $p$ is a Lagrange multiplier.

The equations of motion follow from (\ref{eq:lagrangianform}) by using the principle of least action~\citep{galley2013classical}, giving 
\begin{eqnarray}
  \nu_q \dot{q} & = & - \dpar{F}{q} + p \dpar{V}{q}, \label{eq:general-eqM-q} \\
  \nu_\epsilon \dot{\epsilon} & = & - \dpar{F}{\epsilon} + p \dpar{V}{\epsilon}. \label{eq:general-eqM-eps}
\end{eqnarray}
The terms on left-hand side of~(\ref{eq:general-eqM-q}) and~(\ref{eq:general-eqM-eps}) correspond to the friction forces arising from the motion of the liquid, where $\nu_q$, and $\nu_\epsilon$ are drag coefficients. These can be determined using a Rayleigh dissipation function, defined as~\citep[see][]{goldstein2002classical} 
\begin{equation}
\dot{\mathcal{E}} \equiv \sum_\xi \nu_{\xi} {\dot \xi}^2,
\label{eq:rayleigh}
\end{equation} 
where $\xi = q, \epsilon$. Note that  we have defined $\dot{\mathcal{E}}$ such that the out-of-equilibrium surface energy is dissipated by $\dot{\mathcal{E}}$, i.e., $\dd F / \dd t + \dot{\mathcal{E}} = 0$.
Differentiating~(\ref{eq:rayleigh}) with respect to $\dot \xi$ gives the drag coefficients, 
\begin{equation}
  \nu_{\xi} = \frac{1}{2} \frac{\partial^2 \dot{\mathcal{E}}}{\partial \dot{\xi}^2}.
  \label{eq:dragcoefficientsgeneral}
\end{equation}

An equivalent but more direct approach to describe dynamics is to use~(\ref{eq:energycomposition}) to obtain the total capillary force arising from a change in the position of the barrel, $X$,  and equating this to an effective drag force, i.e.,
\begin{equation}
  \nu_X \dot{X} = - \dtot{F_V}{X}.
  \label{eq:singlevarmotion}
\end{equation}

The drag coefficient $\nu_X$ can be related to $\nu_q$ and $\nu_\epsilon$. Enforcing the conservation of volume explicitly gives
\begin{equation}
  \dot{V} = \dpar{V}{q} \dot{q} + \dpar{V}{\epsilon} \dot{\epsilon} = 0.
  \label{eq:dot_V}
\end{equation}
This relation can then be used in conjunction with (\ref{eq:general-eqM-q}) and~(\ref{eq:general-eqM-eps}) to obtain 
\begin{equation}
  \nu_X = \left( \frac{{\rm d} q}{{\rm d} X}\right)^2 \left\{ \nu_q + \left( \dtot{\epsilon}{q} \right)^2 \nu_\epsilon \right\}.
  \label{eq:dragcoefficientsrelation}
\end{equation}

\subsection{Sources of dissipation}
\label{Estimating the drag coefficient}

As discussed by \citet{degennes1985wetting} and~\citet{ruijter1999droplet}, the total energy dissipation arising  during the motion of a meniscus, $\dot {\mathcal{E}}$, results from three main contributions, 
\begin{equation}
  \dot{\mathcal{E}} = \dot{\mathcal{E}}_\mathcal{H} + \dot{\mathcal{E}}_\mathcal{L} + \dot{\mathcal{E}}_\mathcal{F},
  \label{eq:totalE}
\end{equation}
where $\dot{\mathcal{E}}_\mathcal{H}$ is the hydrodynamic dissipation, $\dot{\mathcal{E}}_\mathcal{L}$ is the dissipation due to the contact line motion, and $\dot{\mathcal{E}}_\mathcal{F}$ is the energy dissipation arising from the formation of a precursor film ahead of the contact line. The latter term is negligible for partial wetting situations, and therefore we shall set $\dot{\mathcal{E}}_\mathcal{F} = 0$.

The hydrodynamic contribution, $\dot{\mathcal{E}}_\mathcal{H}$, is the rate of viscous dissipation caused by the flow pattern. Denoting $\bm{u}$ the velocity field in the liquid, we have~\citep{landau2013fluid},
\begin{equation}
  \dot{\mathcal{E}}_\mathcal{H} = \frac{1}{2} \eta \int_V \left( \bnabla \bm{u} + \bnabla \bm{u}^T \right)^2 \, \dd V. 
  \label{eq:doteh}
\end{equation}
The integral in~(\ref{eq:doteh}) can be split into a `bulk' contribution, arising from the flow pattern at length scales comparable to the barrel size, and a contribution coming from the corner flow 
near the contact line. Therefore, 
\begin{equation}
  \dot{\mathcal{E}}_\mathcal{H} = \dot{\mathcal{E}}_\text{bulk} + \dot{\mathcal{E}}_\text{corner}.
\end{equation}
The bulk dissipation can be estimated by assuming a local Jeffery-Hamel flow~\citep{jeffery1915twodim,hamel1917spiralformige}, which is a pressure-driven flow between two non-parallel planes~\citep{rosenhead1939steady}. In the limit of small $\beta$ we obtain
\begin{equation}
  \dot{\mathcal{E}}_\text{bulk} \approx \frac{12 \upi \eta X^2 W^2}{\beta (4X^2 - W^2)^{3/2}} \dot{X}^2 \approx \frac{6 \upi \eta}{\beta^2} |\cos \theta_\text{e} | (1 + \epsilon)^2 \, q \, \dot{q}^2,
  \label{eq:dissipationbulk}
\end{equation}
where $W \approx 2 q (1 + \epsilon)$ (see Appendix B for details of the calculation of the Jeffery-Hamel flow).

The energy dissipation due to the corner flow near the contact line arises from a deviation of the meniscus from its equilibrium configuration. This can be quantified in terms of a dynamic contact angle, $\theta$. The capillary force driving the distortion of the interface is 
\begin{equation}
  \bm{f}_\text{cl} = \gamma (\cos \theta_\text{e} - \cos \theta) \hat{\bm{r}}_\text{cl} \approx \gamma \sin \theta_\text{e} (\theta - \theta_\text{e}) \hat{\bm{r}}_\text{cl},
  \label{eq:force-cl}
\end{equation}
where $\hat{\bm{r}}_\text{cl}$ is the unit normal of the contact line. Therefore, the energy dissipation can be expressed as~\citep[see, e.g.,][]{degennes1985wetting},
\begin{equation}
  \dot{\mathcal{E}}_\text{corner} = 2 \int_0^{2\upi} \bm{f}_\text{cl} \bcdot \bm{v}_{\rm cl} \, r_\text{cl} \,\dd \varphi,
  \label{eq:cvdissipation}
\end{equation}
where $\bm{v}_{\rm cl} = \dot{\bm{x}}_\text{lg}(\varphi,\vartheta = \psi)$ and $r_\text{cl} = r_\text{cl}(\varphi)$ are the velocity and radius the contact line, respectively. 

The deviation of the dynamic contact angle from the equilibrium value can be estimated using the Cox-Voinov law~\citep{voinov1976hydrodynamics,cox1986dynamics1}, 
\begin{equation}
  \frac{\theta_\text{e} - \cos \theta_\text{e} \sin \theta_\text{e}}{2 \cos \theta_\text{e}} (\theta - \theta_\text{e}) = \frac{\eta}{\gamma} v_\text{cl} \log \frac{ \ell_{\rm M}}{\ell_\text{m}}, 
  \label{eq:coxvoinov}
\end{equation}
where $v_\text{cl} =  \bm{v}_{\rm cl} \bcdot  \hat{\bm{r}}_\text{cl}$. 
The logarithmic factor in~(\ref{eq:coxvoinov}) quantifies the 
relative contribution to the dissipation from the corner flow at the macroscopic length scale, $\ell_{\rm M} \sim q$, and the microscopic lengthscale, $\ell_\text{m}$.  The microscopic length scale acts as a cut-off to regularise the viscous dissipation singularity~\citep{huh1971hydrodynamic} and depends on the details of the liquid-gas interactions and the roughness of the solid surface~\citep{bocquet2010nanofluidics}. 

Using~(\ref{eq:coxvoinov}) to eliminate $\theta - \theta_\text{e}$ from~(\ref{eq:cvdissipation}) gives 
\begin{equation}
  \dot{\mathcal{E}}_\text{corner} = \frac{4 \eta \sin^2 \theta_\text{e}}{\theta_\text{e} - \cos \theta_\text{e} \sin \theta_\text{e}} \log \frac{q}{\ell_\text{m}} \int_0^{2 \upi} v_\text{cl}^2 r_\text{cl} \, \dd \varphi,
  \label{eq:dissipationcoxvoinov}
\end{equation}
where, to leading order in $\beta$,
\begin{equation}
  \int_0^{2 \upi} v_\text{cl}^2 r_\text{cl} \, \dd \varphi = \frac{\upi}{\beta^2} \cos^2 \theta_\text{e} | \sin \theta_\text{e} + \epsilon | \, q \, \dot{q}^2 + O(\beta^0).
  \label{eq:contact-line-integral-app}
\end{equation}

At length scales smaller than $\ell_{\rm m}$, the dissipation is controlled by the motion of the liquid and gas molecules at the contact line. Matching the average speed of the molecules to $v_{\rm cl}$ one obtains the friction law~\citep{ruijter1999droplet},
\begin{equation}
  \dot{\mathcal{E}}_\mathcal{L} = 2 \zeta_0 \int_0^{2 \upi} v_\text{cl}^2 r_\text{cl} \, \dd \varphi,
  \label{eq:dissipationcontactline}
\end{equation}
where the friction coefficient, $\zeta_0$, is determined by the competition between the adsorption of molecules by the solid and thermal fluctuations~\citep{blake1969kinetics}. 

\subsection{Relaxation towards equilibrium}
\label{subsection:Relaxation towards equilibrium}

Close to equilibrium, the restoring force in~(\ref{eq:singlevarmotion}) can be obtained by expanding $F_V$ in powers of $X - X_{\rm e}$, i.e., 
\begin{equation}
  F_V(X) = F_\text{e} + \frac{1}{2} k (X - X_\text{e})^2 + O(X - X_\text{e})^3,
  \label{eq:singlevarenergy}
\end{equation}
where the coefficient of restitution is   
\begin{equation}
  k = 6 \gamma a_0 \left( 1 - \frac{3 a_0 a_2}{a_1^2} \right) \frac{\sin^2 \beta}{\cos^2 \theta_\text{e}}.
  \label{eq:restitution_constant}
\end{equation}

To estimate the drag coefficient, $\nu_X$, we first substitute~(\ref{eq:dissipationbulk}),~(\ref{eq:dissipationcoxvoinov}) and~(\ref{eq:dissipationcontactline}) into~(\ref{eq:totalE}). Then, using ~(\ref{eq:dragcoefficientsgeneral}), gives, to leading order in $\beta$, 
\begin{eqnarray}
  \nu_q & = &  \frac{2 \upi \eta}{\beta^2} q | \cos \theta_\text{e} | \left[ 3(1 + \epsilon)^2 + \left( \frac{2 \sin^2 \theta_\text{e} \log q / \ell_\text{m}}{\theta_\text{e} - \cos \theta_\text{e} \sin \theta_\text{e}}  + \frac{\zeta_0}{\eta} \right) |\cos \theta_\text{e}| | \sin \theta_\text{e} + \epsilon | \right], \\
  \nu_\epsilon & = & 0.
\end{eqnarray}
The close-to-equilibrium behaviour of $\nu_X$ can then be obtained by substituting $\nu_q$ and $\nu_\epsilon$ into~(\ref{eq:dragcoefficientsrelation}), and setting $q \approx R_{\rm e} = (V / a_0)^{1/3}$ and $\epsilon \approx 0$. This produces the result
\begin{equation}
  \nu_X \approx \frac{6 \upi \eta}{|\cos \theta_{\rm e}|} \left[ 1 + {\sin \theta_\text{e} |\cos \theta_\text{e}|} \left( \frac{2 \sin^2 \theta_\text{e} \log (R_{\rm e}/ \ell_\text{m})}{3(\theta_\text{e} - \cos \theta_\text{e} \sin \theta_\text{e})}   + \frac{\zeta_0}{3\eta} \right) \right] \left( \frac{V}{a_0} \right)^{1/3}.
  \label{eq:netdragcoefficient}
\end{equation}

This expression gives the relative contributions to the drag coefficient arising from the bulk, corner and contact-line, corresponding to the first, second and third terms inside the square brackets, respectively. 


Using~(\ref{eq:netdragcoefficient}) and~(\ref{eq:singlevarenergy}) in~(\ref{eq:singlevarmotion}) gives the exponential relaxation of the position of the barrel towards equilibrium,
\begin{equation}
  X(t) = X_\text{e} + (X(0) - X_\text{e}) {\rm e}^{- t / \tau},
  \label{eq:time-evolution-droplet-centre}
\end{equation}
where the ratio 
\begin{equation}
\tau \equiv \frac{\nu_X}{ k} 
\end{equation}
sets the time scale of the relaxation process.


\section{Discussion and Conclusions}
\label{section:Discussion and Conclusions}

\citet{concus2001bridges} predicted the existence of equilibrium barrel shapes, which correspond to sections of a sphere. Such states exist so far as the contact and wedge angles satisfy $90^\circ + \beta  < \theta_{\rm e} < 180^\circ$. 

\begin{figure}
  \centering
  \begin{subfigure}{\textwidth}
   \includegraphics[width=\textwidth]{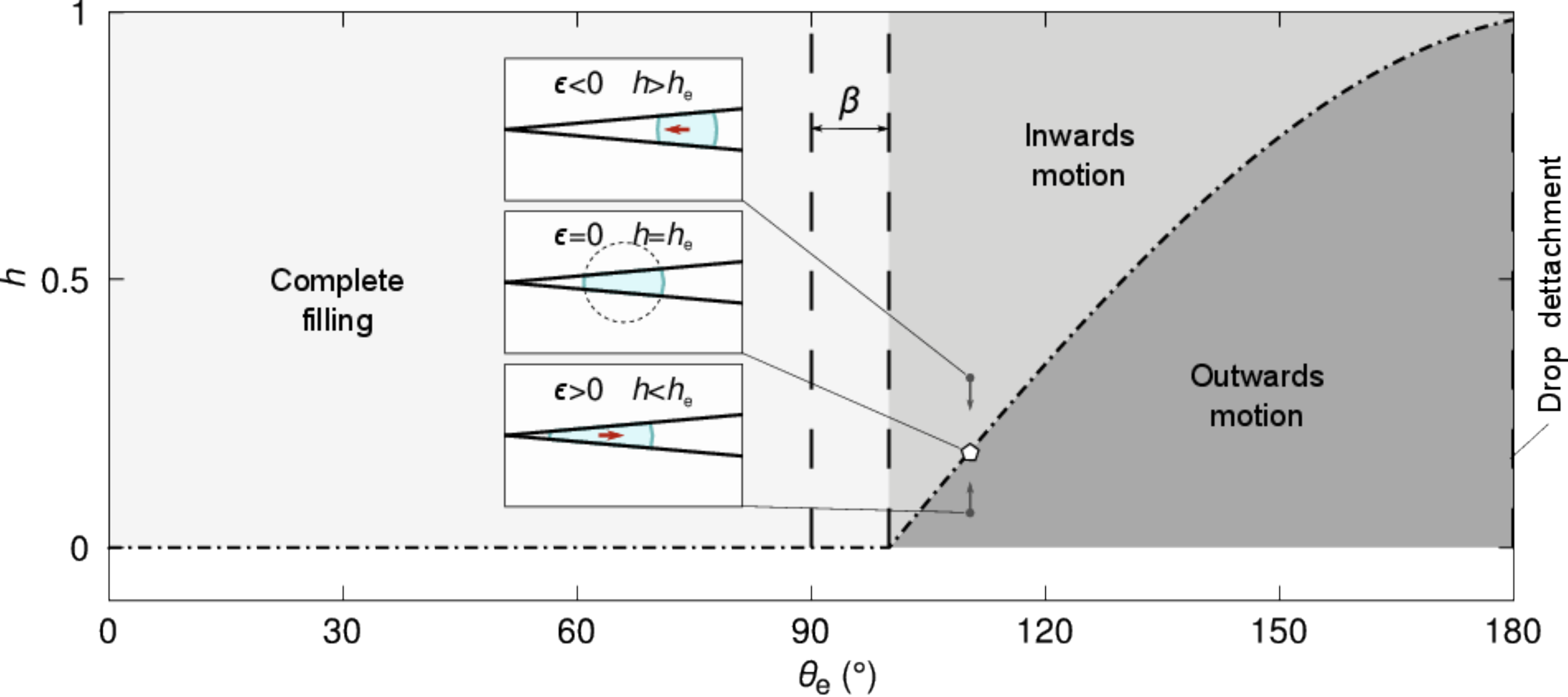}
  \end{subfigure}
  \caption{ \label{fig:h_plot} (Colour online) Phase diagram of the filling states of liquid droplet in a solid wedge. The vertical arrows indicate trajectories of the system for fixed values of $\theta_\text{e}$ and $\beta$. The dashed line shows the onset of edge blobs. The equilibrium of the trajectories is highlighted and examples of the morphology of the droplet are shown as insets. For clarity, the volume in the examples is not the same. }
\end{figure}

In equilibrium, the height-to-width aspect ratio of the barrel, $h_\text{e}$, plays the role of an order parameter. This idea is illustrated figure~\ref{fig:h_plot}, which shows a phase diagram of the different filling regimes for a wedge. For $\theta \leq 90^\circ + \beta$, $h_\text{e} = 0$, corresponding to the complete filling states studied by~\citet{concus1998discontinuous}, \citet{concus2001bridges} and~\citet{brinkmann2004}. For $\theta_\text{e} > 90^\circ + \beta$, corresponding to the barrel regime, the aspect ratio becomes finite, i.e.,  $h_\text{e} = - \cos(\theta_\text{e} - \beta)$. Increasing the equilibrium contact angle leads to a limiting barrel configuration, where $\theta_\text{e} = 180^\circ$ and  $h_\text{e} = \cos \beta$. In such a limit the contact area between the liquid and the solid vanishes, and the liquid forms a suspended droplet. 

Out of equilibrium, the instantaneous aspect ratio characterises the inwards and outwards modes of motion for a liquid barrel. Using~(\ref{eq:aspectratio}) and~(\ref{eq:rq}) gives 
\begin{equation}
 h = h_\text{e} \, \frac{\cos \theta_\text{e} + (1 + \epsilon) \sin \beta}{(1 + \epsilon) (\cos \theta_\text{e} + \sin \beta)}.
 \label{eq:hnoneq}
\end{equation}
A displacement of the liquid towards the apex of the wedge will result in a vertical compression of the barrel (see upper inset in figure~\ref{fig:h_plot}). This corresponds to setting $\epsilon > 0$ in (\ref{eq:hnoneq}), which leads to a decrease in the aspect ratio, i.e., $h < h_\text{e}$. In contrast, a displacement towards the wide end of the wedge causes a vertical extension of the interface, and corresponds to $\epsilon < 0 $, or, equivalently, $h > h_\text{e}$. The energy landscapes reported in~\S\ref{subsection:Near equilibrium} suggest that the spherical barrel shapes correspond to global minima in the surface energy, and therefore distortions to such shapes will always relax back to equilibrium. For situations where $\theta_\text{e}$ and $\beta$ are kept constant, trajectories towards equilibrium run as vertical lines in figure~\ref{fig:h_plot}, pointing towards the master curve $h_\text{e} (\theta_\text{e} - \beta)$.

\citet{baratian2015shape} showed that the equilibrium barrel shapes are subject to a vanishing net force. In such a case, the pressure force exerted by the walls on the droplet is exactly balanced by the surface tension acting along the contact line. Out of equilibrium, the net force will not vanish. 
Because the mean curvature of the barrels is positive, the average Laplace pressure within the droplet is larger than the pressure of the surrounding medium. 
Therefore, the lateral projection of the pressure force exerted by the liquid on the walls of the wedge points towards the apex, and, consequently, the walls will always exert a reaction force pointing in the outwards direction.
On the other hand, the tension acting over the contact line can always be locally decomposed into components that are normal and tangential to the solid surface. In situations where the contact angle is uniform, the integral of the tangential component of the tension force over the contact line will vanish. The vertical component will always point towards the solid, and therefore, the walls will exert a net force pointing inwards.  
Because a displacement of the barrel towards the narrow end of the wedge always results in an outwards motion, one can infer that the pressure force must be larger than the tension force. On the other hand, the inwards motion of the barrels from the wide end of the wedge towards their equilibrium position suggests that the dynamics is dominated by the tension acting at the contact line. 
These features are analogous to the well-known problem of capillary invasion~\citep{degennes2013capillarity}. 
For tension-dominated dynamics one expects little deviation of the local contact angle from the equilibrium contact angle, similarly to the spontaneous imbibition problem, whilst in the pressure-dominated regime the interface should deform more appreciably from its equilibrium configuration, as in forced imbibition.  

\begin{figure}
  \centering
  \includegraphics[width=\textwidth]{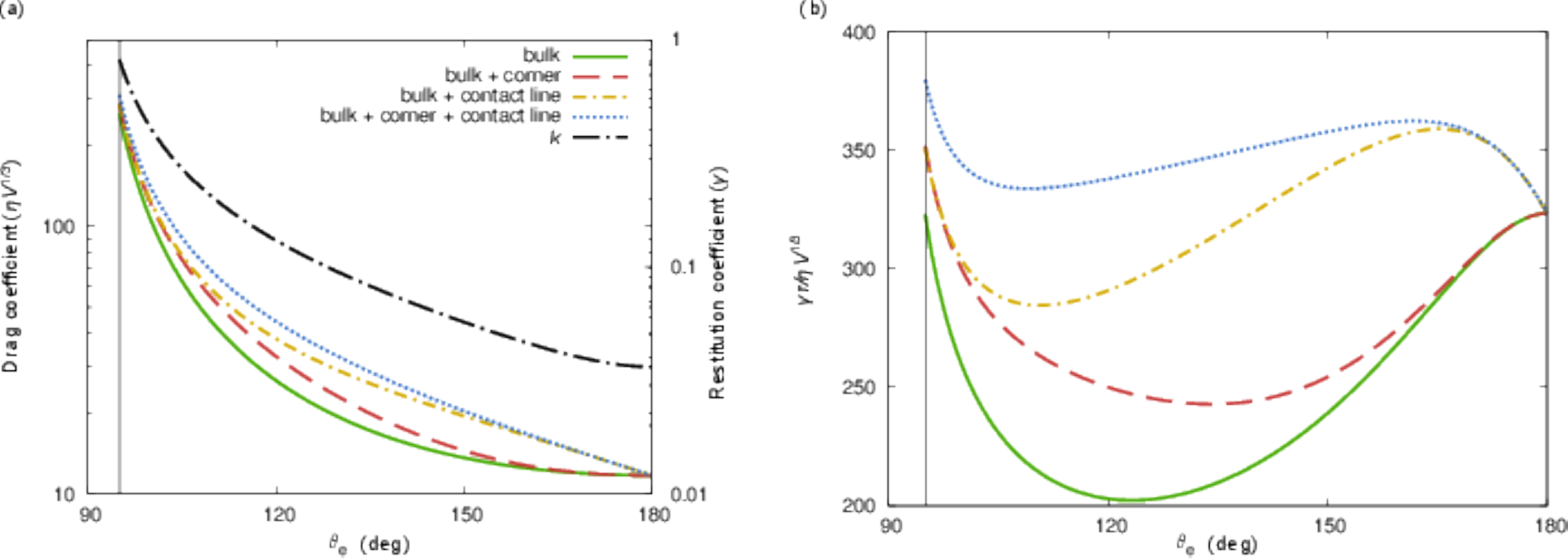}
  \caption{ \label{fig:relaxation_time} (Colour online) Bulk, corner flow and contact line contributions to (a) the drag coefficient, $\nu_X$ and (b) the relaxation time, $\tau$, of the translational motion of barrels along  the bisector plane of a wedge of angle $\beta = 5^\circ$. In (a) the restitution coefficient, $k$, is superimposed on the right-hand side axis. The vertical lines in both plots correspond to the limiting wetting angle $\theta_{\rm e} = 90^\circ + \beta.$  
}
\end{figure}

We close our discussion by focusing on the relaxation time scale of the translational motion of the barrels, $\tau=\nu_X/k$, which results from the balance of the driving capillary force, characterised by the restitution constant $k$, and the overall friction, characterised by the drag coefficient, $\nu_X$. 
Figures~\ref{fig:relaxation_time}(a) and (b) show plots of $k$, $\nu_X$, and $\tau$ as functions of the equilibrium angle. 
In the limit $\theta_{\rm e}\rightarrow 90^\circ + \beta$, the barrel equilibrium position is closer to the apex of the wedge. Geometrically, this implies a stronger confinement, and thus both the bulk contribution to the friction coefficient and the restitution constant reach local maxima in this limit. 
For larger $\theta_{\rm e}$, both quantities decrease monotonically, leading to an initial decrease in the relaxation time.  However, the rate at which $k$ decreases becomes dominant with increasing $\theta_{\rm e}$. This is because at higher equilibrium contact angle the barrels keep an approximately spherical shape for larger displacements from equilibrium. As a result, the relaxation time   reaches a minimum, beyond which it increases with $\theta_{\rm e}$ until it reaches a maximum saturation value as $\theta_{\rm e}\rightarrow 180^\circ$.  
Interestingly, the location of the minimum and maximum relaxation times shown in figure~\ref{fig:relaxation_time}(b) depends on the contributions to dissipation from the motion of the corner flow and the contact line.  

The typical magnitude of the corner flow is controlled by the length scale separation between the macroscopic length scale $\ell_{\rm M}\sim R_{\rm e}$, and the microscopic length scale $\ell_{\rm m}$ that characterises the molecular processes at the level of the contact line~\citep{Snoeijer2013}. For a macroscopic droplet, $R_{\rm e}\sim 1~{\rm mm}$ and $\ell_{\rm m}\sim 10~{\rm nm}$, and thus $R_{\rm e}/\ell_{\rm m}\approx 10^5$. As shown in figure~\ref{fig:relaxation_time}(a), this additional contribution is important at intermediate angles, and vanishes in the limit $\theta_{\rm e} \rightarrow 180^\circ$. This is the combined effect of a vanishing contour length and a less confined corner flow at higher opening angle. As a result of the corner flow,  the minimum in the relaxation time is displaced  to a higher contact angle, as shown in figure~\ref{fig:relaxation_time}(b).   
 
The contribution of contact line dissipation to the drag coefficient is controlled by the (constant) microscopic friction coefficient $\zeta_0$ and the contour length of the contact line. Therefore, this term decays more slowly than the corner flow term in~(\ref{eq:netdragcoefficient}). The relative weight of these sources of dissipation, however, is controlled by the ratio $\zeta_{0}/\eta$. Estimating $\zeta_0$ will in general be subject to the details of a specific model~\citep[see, e.g.,][]{ranabothu2005dynamic,sikalo2005dynamic}. Rather,  here we examine the case where $\zeta_{0}/3\eta = 1$ in~(\ref{eq:netdragcoefficient}) as a specific example where the corner and contact line dissipation are comparable in magnitude. As shown in figures~\ref{fig:relaxation_time}(a) and~\ref{fig:relaxation_time}(b), the main effect of this term is a slower decay in the contact line dissipation with increasing contact angle, which in turn leads to an overall broadening of the maximum in the relaxation time. 

Therefore, the qualitative shape for $\tau$ {\it vs} $\theta_{\rm e}$ can be used in experiments to identify the relative contribution of each source of dissipation in the motion of the liquid barrels. more quantitatively, our model can be used to estimate the values of the microscopic cut-off length, $\ell_\text{m}$, and the friction coefficient, $\zeta_0$, by treating these quantities as fitting parameters.


\begin{acknowledgements}
We would like to thank G. G. Wells and J. Guan for bringing to our attention their interesting experimental observations and discussions which motivated this work, and to F. Mugele for enlightening discussions.
\'E. Ruiz-Guti\'errez is supported by a Northumbria University PhD Studentship.  C. Semprebon acknowledges support from Northumbria University through the Vice-Chancellor's Fellowship Programme. 
\end{acknowledgements}

\appendix

\section{Coefficients of the volume and free energy polynomial forms}
\label{app:Volume-polynomial-form-coefficients}

Integrating (\ref{eq:xlg}) with respect to $\vartheta$ gives the following expression for the volume of the liquid,
\begin{equation}
  \begin{split}
    V = \frac{1}{3} \int_0^{2 \upi} & R \left\{2 (X \partial_\varphi r \sin \varphi + r^2 + X r \cos \varphi) \sin \psi + R [r (3 \psi +\sin \psi \cos \psi) \right. \\
         & \left. + X (\psi + \sin \psi \cos \psi) \cos \varphi ] + 2 X \partial_\varphi R \psi \sin \varphi + 2 R^2 \sin \psi \right\} \, \dd \varphi,
  \end{split}
  \label{eq:vatheta}
\end{equation}

Following the approximations detailed in \S\ref{section:Morphology of the droplet}, the radii $r$ and $R$ read
\begin{eqnarray}
  r(\varphi) & = & \frac{q \epsilon \alpha}{\cos \varphi + \alpha}, \label{eq:r_phi} \\
  R(\varphi) & = & q \left( 1 + \frac{ \epsilon \cos \varphi }{\cos \varphi + \alpha } \right), \label{eq:R_phi}
\end{eqnarray}
where $\alpha = - \cos \theta_\text{e} / \sin \beta$.

Substituting (\ref{eq:r_phi}) and (\ref{eq:R_phi}) into (\ref{eq:vatheta}) results in a polynomial of $\epsilon$ 
where the coefficients are,
\begin{equation}
  a_0 = \frac{1}{3} \int_0^{2\upi} \left\{ 2 \sin \psi + \alpha [ \psi +\sin \psi \cos \psi ] \cos \varphi \right\} \, \dd \varphi,
  \label{eq:expr_a0}
\end{equation}
\begin{equation}
  \begin{split}
    a_1 =  \frac{1}{6} \int_0^{2 \upi} (\cos \varphi + \alpha)^{-3} & \left\{ 4 \alpha ^4 \cos \varphi \sin \psi \right. \\
    & + \alpha^3 [ (4 \psi + 2 \sin \psi + \sin 2 \psi ) \cos 2 \varphi + 6 (\psi +\sin \psi )+2 \sin 2 \psi ] \\
    & + 2 \alpha ^2 [ ( \{3 \psi + \sin 2 \psi \} \cos 2 \varphi + 7 \psi + 4 \{ 2 + \cos \psi\} \sin \psi ) \cos \varphi ] \\
    & + \alpha [ (2 \psi +\sin 2 \psi ) \cos 2 \varphi + 8 \psi + 4 (6 + \cos \psi) \sin \psi ] \cos^2 \varphi \\
    & \left. + 12 \cos^3 \varphi \sin \psi \right\} \, \dd \varphi,
  \end{split}
  \label{eq:expr_a1}
\end{equation}
\begin{equation}
  \begin{split}
    a_2 = \frac{1}{12} \int_0^{2 \upi} (\cos \varphi + \alpha)^{-3} & \left\{ 4 \alpha ^3 [\cos 2 \varphi + 3] \sin \psi \right. \\
    & + \alpha^2 [ (6 \psi +\sin 2 \psi) \cos 2 \varphi + 22 \psi +16 \sin \psi + 5 \sin 2 \psi ] \cos \varphi \\
    & + \alpha [ (2 \psi +\sin 2 \psi) \cos 2 \varphi + 26 \psi +24 \sin \psi + 5 \sin 2 \psi ] \cos^2 \varphi \\
    & \left. + 24 \cos^3 \varphi \sin \psi \right\} \, \dd \varphi,
  \end{split}
  \label{eq:expr_a2}
\end{equation}
and
\begin{equation}
  \begin{split}
    a_3 = \frac{1}{6} \int_0^{2 \upi} (\cos \varphi + \alpha)^{-3} & \left\{ 4 \alpha ^2 \cos \varphi \sin \psi \right. \\
    & +\alpha  \left[ 6 \psi + \sin 2 \psi \right] \cos^2 \varphi \\
    & \left. + 4 \cos^3 \varphi \sin \psi \right\} \, \dd \varphi.
  \end{split}
  \label{eq:expr_a3}
\end{equation}

The evaluation of the integrals requires an explicit expression of the angle $\psi$; using (\ref{eq:contactangleapprox}) gives
\begin{equation}
  \psi = \arctan \left[ \frac{\tan \beta}{1 + \alpha^2 \tan^2 \beta} \left( \alpha \sqrt{1 + (\cos^2 \varphi - \alpha^2 ) \tan^2 \beta} + \cos \varphi \right) \right].
  \label{eq:long_psi}
\end{equation}
Using~(\ref{eq:long_psi}) and expanding in powers of $\beta$ leads to~(\ref{eq:vol_coefficients}).

To compute the interfacial energy we first note that~(\ref{eq:freeenergygeneral}) is composed of two terms, the first being
\begin{equation}
  \int_0^{2 \upi} \int_{-\psi}^{\psi} | \dd \bm{A}_\text{lg} | = \int_0^{2 \upi} \int_{-\psi}^\psi R \left[ R^2 \cos^2 \vartheta + 2 R r \cos \vartheta + r^2 + (\partial_\varphi R + \partial_\varphi r \cos \vartheta)^2\right]^{1/2} \, \dd \vartheta \dd \varphi.
\end{equation}
The integral in $\vartheta$ can be expressed in terms of elliptic functions. Then, substituting $R$ and $r$ using~(\ref{eq:r_phi}) and (\ref{eq:R_phi}), gives an expression in terms of $q$ and $\epsilon$. Close to equilibrium $\epsilon\ll1$, therefore, we evaluate the integral by first expanding the integrand in powers of $\epsilon$, which leads to~(\ref{eq:epoly}).


\section{Jeffery-Hamel flow}
\label{app:Jeffery-Hamel-flow}

The Jeffery-Hamel flow is a two-dimensional, pressure driven, radial, steady flow contained inside flat walls diverging at an angle $2\beta$.

In the low Reynolds number regime, the fluid dynamics can be described using the Stokes equation~\citep{pozrikidis1992boundary},
\begin{equation}
  \eta \nabla^2 \bm{u} - \nabla P = 0,
  \label{eq:Stokes-flow}
\end{equation}
where $P$ is the pressure field and $\bm{u}$ is the velocity field. For an incompressible fluid, the continuity equation reduces to 
\begin{equation}
  \bm{\nabla} \bcdot \bm{u} = 0.
  \label{eq:continuity}
\end{equation}

Using the polar coordinates ($s$, $\omega$) the velocity field is expressed as $\bm{u} = u_s \hat{\bm{s}} + u_\omega \hat{\bm{\omega}}$ where the angular flow is assumed to be zero, i.e., $u_\omega = 0$. 

The continuity equation reads,
\begin{equation}
  \frac{1}{s} \dpar{s u_s}{s} = 0,
\end{equation}
which has the general solution,
\begin{equation}
  u_s(s, \omega) = \frac{f(\omega)}{s}
\end{equation}
where $f$ depends only on $\omega$. The explicit form of $f$ can be found using the Stokes equation. We write
\begin{eqnarray}
  \eta \frac{f''}{s^3} - \dpar{P}{s} & = & 0, \qquad \label{eq:Stokes-s} \\
  2 \eta \frac{f'}{s^3} - \frac{1}{s} \dpar{P}{\omega} & = & 0. \label{eq:Stokes_th}
\end{eqnarray}
Integrating (\ref{eq:Stokes_th}) with respect to $\omega$ gives the pressure profile,
\begin{equation}
  P(s, \omega) = \frac{2 \eta}{s^2} f(\omega) + g(s),
\end{equation}
where $g$ only depends on $s$. Substituting this result into (\ref{eq:Stokes-s}) gives the equation,
\begin{equation}
  \dtot{^2 f}{\omega^2} + 4 f = \frac{s^3}{\eta} \dtot{g}{s}.
\end{equation}
The left hand side only depends on $\omega$, whereas the right hand side only depends on $s$, the only way that this can happen is if both sides are equal to a constant, $c_1$, then,
\begin{equation}
  g(s) = - \frac{c_1 \eta}{2 s^2} + c_2, \qquad \text{and} \qquad f(\omega) = \frac{c_1}{4} + c_3 \cos 2 \omega + c_4 \sin 2 \omega.
\end{equation}
The constants $c_i$, $i = 1, ..., 4$ can be found by imposing boundary conditions to the flow. Due to symmetry, the flow profile must be an even function of $\omega$, therefore $c_4 = 0$. Imposing a no-slip boundary condition at the walls of the wedge fixes $c_1 = -4 c_3 \cos 2 \beta$. Setting the pressure to $P(s_1)=P_1$ and $P(s_2)=P_2$ at two arbitrary points, $s_1$ and $s_2$, fixes $c_2 = (P_1 s_1^2 - P_2 s_2^2) / (s_1^2 - s_2^2)$ and $c_3 = (P_2 - P_1) s_1^2 s_2^2 / 2 \eta (s_1^2 - s_2^2)$. 

The average velocity of the Jeffery-Hamel flow  over a region of the channel reads 
\begin{eqnarray}
  \dot{X} & = & \left[ \int_{-\beta}^{\beta} \int_{s_1}^{s_2} s \dd s \dd \omega \right]^{-1} \int_{-\beta}^{\beta} \int_{s_1}^{s_2} \bm{u} \bcdot \hat{\bm{s}} \, s \dd s \dd \omega \\
  & = & - \frac{P_1 - P_2}{s_1 - s_2} \left( \frac{s_1 s_2}{s_1 + s_2} \right)^2 \frac{\sin 2 \beta - 2 \beta \cos 2 \beta}{2 \beta \eta}.
\end{eqnarray}
In terms of $\dot{X}$, the velocity field reads,
\begin{equation}
  \bm{u} = \dot{X} \frac{s_1 + s_2}{2 s} \frac{\cos 2 \beta - \cos 2 \omega}{\cos 2 \beta - \beta^{-1} \sin 2 \beta } \hat{\bm{s}}. 
  \label{eq:velfieldxdot}
\end{equation}

Equation~(\ref{eq:velfieldxdot}) can be used to determine the bulk energy dissipation. First, the gradient of the velocity field is
\begin{equation}
  \bm{\nabla} \bm{u} = \frac{\dot{X} \beta (s_1 + s_2)}{s^2 (2 \beta \cos 2 \beta - \sin 2 \beta)} \left[
  \begin{matrix}
    \cos 2 \omega - \cos 2 \beta & \sin 2 \omega \\
    0 & \cos 2 \beta - \cos 2 \omega \\
  \end{matrix}
  \right],
\end{equation}
which leads to bulk energy dissipation density 
\begin{equation}
  \dot{\varepsilon} = \frac{\eta}{2} (\bm{\nabla} \bm{u} + \bm{\nabla} \bm{u}^T)^2 = \frac{2 \eta \dot{X}^2 \beta^2 (s_1 + s_2)^2 (3 + \cos 4 \beta - 4 \cos 2 \beta \cos 2 \omega)}{s^4 (2 \beta \cos 2 \beta - \sin 2 \beta)^2}.
  \label{eq:bedd}
\end{equation}
To obtain the total dissipation,~(\ref{eq:bedd}) needs to be integrated over a volume $V_\text{eff}$, which corresponds to the region where the bulk dissipation of the barrel takes place. 
We approximate $V_\text{eff}$, as a toroidal section, of major radius equal to the distance $X$, and a minor radius that matches the equatorial radius of the barrel, $W/2$. Therefore, the bulk dissipation is
\begin{equation}
  \dot{\mathcal{E}}_\text{bulk} = \int \dot{\varepsilon} \, \dd V_\text{eff} \approx \frac{32 \pi \beta ^2 \eta \dot{X}^2 W^2 X^2 [ \beta (\cos 4 \beta + 3) - \sin 4 \beta]}{(4 X^2 - W^2)^{3/2} (2 \beta  \cos 2 \beta - \sin 2 \beta)^2}.
  \label{eq:bulk_dissipaiton_jh}
\end{equation}
To a leading order, (\ref{eq:bulk_dissipaiton_jh}) is inversely proportional to $\beta$, which leads to the expression (\ref{eq:dissipationbulk}) after taking a Laurent series expansion.



\end{document}